\documentstyle[amsfonts,aps,eqsecnum]{revtex}

\begin{document}
\draft
\title{Thermodynamics of Coupled Identical Oscillators within the Path Integral
Formalism}
\author{F.~Brosens and J.T. Devreese\cite{Author2},}
\address{Departement Natuurkunde, Universiteit Antwerpen (UIA),\\
Universiteitsplein 1, B-2610 Antwerpen}
\author{L. F. Lemmens,}
\address{Departement Natuurkunde, Universiteit Antwerpen (RUCA),\\
Groenenborgerlaan 171, B-2020 Antwerpen}
\date{September 9, 1996}
\maketitle

\begin{abstract}
A generalization of symmetrized density matrices in combination with the
technique of generating functions allows to calculate the partition function
of identical particles in a parabolic confining well. Harmonic two-body
interactions (repulsive or attractive) are taken into account. Also the
influence of a homogeneous magnetic field, introducing anisotropy in the
model, is examined. Although the theory is developed for fermions and
bosons, special attention is payed to the thermodynamic properties of bosons
and their condensation.
\end{abstract}


\section{Introduction}

The study of the density matrices of identical particles (bosons or
fermions) moving freely in a box \cite{FeynStat} is generalized in this
paper to the case of identical particles in a parabolic confinement
potential with either harmonic interactions between the particles, or with
an anisotropy induced by a homogeneous magnetic field on top of the
parabolic confinement. This model, giving rise to repetitive Gaussian
integrals, allows to derive an analytical expression for the generating
function of the partition function. For an ideal gas of non-interacting
particles in a parabolic well, this generating function coincides with the
grand-canonical partition function. With interactions, the calculation of
this generating function circumvents the constraints on the summation over
the cycles of the permutation group. Moreover, it allows one to calculate
the canonical partition function recursively for the system with harmonic
two-body interactions. The theory is developed both for fermions and for
bosons. In view of the recent interest in Bose-Einstein condensation in a
trap \cite{BEC1,BEC2,BEC3}, more attention has been payed to the boson case
in the discussion of the results. The model system discussed here has
already been studied in the context of quantum dots with operator
techniques, and the eigenvalues and eigenstates were calculated including
the effect of harmonic two-body interactions and in the presence of a
magnetic field \cite{Johnson}. However, to the best of our knowledge neither
the boson case nor the thermodynamics seems to have been analyzed
previously. It should also be mentioned that the idea of first expanding the
Hilbert space to the configuration space and then projecting onto the
appropriate subspace by group theoretical means has been used recently \cite
{Haase,Hausler} in the context of quantum dots to study the ground state
correlations for fermions and bosons.

Another motivation to perform the present analytical calculations lies in
our path integral formulation of the density matrices for $N$ particles\cite
{LBD,BDL,LBDPR}. Indeed, particles in a parabolic potential are a favorite
testing ground for the path integral method\cite{FeynHibbs,FeynKlei,GiachTog}%
. It should be noted that for our formulation the existence of a positive
measure over a well defined domain in the $R^{3N}$ configuration space is
essential in view of any algorithmic approach to the problem. In the present
paper, which is in essence analytical, integrations over the configuration
space are performed. The reason is that the extension of the state space to
the configuration space makes the Gaussian integrals tractable. The
permutation symmetry leads to summations over the cycles which are performed
using the generating function technique, which is one of the main results of
the present paper.

The model of $N$ identical particles in a parabolic well, in the presence of
a magnetic field and with harmonic repulsive or attractive two-body
interactions has its intrinsic value, since it constitutes an exactly
soluble idealization of atoms in a magnetic trap. It should be stressed that
the association of identical particles with each three oscillator degrees of
freedom makes the model $3D$. Without Bose-Einstein or Fermi-Dirac
statistics, i.e. for ``distinguishable'' particles, the model is equivalent
with $3N$ one dimensional oscillators, because each degree of freedom
decouples in such a way that there is no difference in statistical behavior
between $3N$ $1D$ oscillators and $N$ $3D$ oscillators \cite{Ford}.

This paper is organized as follows: the calculation technique is explained
in the next section. In the subsequent section we repeat the same
calculation for the model with a homogeneous magnetic field. In section IV
the results for $1D$ bosons and fermions are given. In section V bosons in $%
3D$ are analyzed in some detail, and in the last section the conclusions are
given.

\section{Harmonically interacting identical particles in a parabolic well.}

In this section we calculate the partition function of $N$ identical
particles with the following Lagrangian including one-body and two-body
potentials 
\begin{equation}
L=\frac 12\sum_{j=1}^N{\bf \dot{r}}_j^2-V_1-V_2;\quad V_1=\frac{\Omega ^2}2%
\sum_{j=1}^N\left. {\bf r}_j\right. ^2;\quad V_2=-\frac{\omega ^2}4%
\sum_{j,l=1}^N\left( {\bf r}_j-{\bf r}_l\right) ^2.
\end{equation}

(Atomic units are used.) The potentials can be rewritten in terms of the
center-of-mass coordinate ${\bf R}$ and coordinates ${\bf u}_j$ describing
the coordinates of the particles measured from the center of mass 
\begin{equation}
{\bf R}=\frac 1N\sum_{j=1}^N{\bf r}_j;\quad {\bf u}_j={\bf r}_j-{\bf R},
\end{equation}
from which 
\begin{equation}
V_1+V_2=V_{CM}+V;\quad V_{CM}=\frac 12N\Omega ^2{\bf R}^2;\quad
V=w^2\sum_{j=1}^N\left. {\bf u}_j\right. ^2,
\end{equation}
with 
\begin{equation}
w=\sqrt{\Omega ^2-N\omega ^2}.  \label{eqeigenw}
\end{equation}
The requirement that $w$ has to be positive expresses the stability
condition that the confining potential has to be strong enough to overcome
the repulsion between the particles. If an harmonic interparticle {\bf %
attraction} is considered, the eigenfrequency $w$ would become $w=\sqrt{%
\Omega ^2+N\omega ^2},$ and no stability condition has to be imposed on the
confining potential. Notice that these transformations do neither
diagonalize the Lagrangian nor the Hamiltonian, because the coordinates $%
{\bf u}_j$ are {\sl not} independent of the center-of-mass coordinate.

Since the system consists in each direction of 1 degree of freedom with
frequency $\Omega $ and $\left( N-1\right) $ degrees of freedom with
frequency $w,$ the propagator 
\begin{equation}
K_D\left( {\bf r}_1^{\prime \prime }\cdots {\bf r}_N^{\prime \prime },\beta |%
{\bf r}_1^{\prime }\cdots {\bf r}_N^{\prime },0\right) \equiv \left\langle 
{\bf r}_1^{\prime \prime }\cdots {\bf r}_N^{\prime \prime }\left| e^{-\beta
H}\right| {\bf r}_1^{\prime }\cdots {\bf r}_N^{\prime }\right\rangle _D
\label{eqKDdef}
\end{equation}
for {\bf distinguishable} particles (indicated by the subscript $D$ for 3
dimensions and $d$ in 1 dimension) can be calculated from the action
expressed in the imaginary time variable, and it is of course a product of
the propagators $K_d$ per component: 
\begin{equation}
K_D\left( {\bf r}_1^{\prime \prime }\cdots {\bf r}_N^{\prime \prime },\beta |%
{\bf r}_1^{\prime }\cdots {\bf r}_N^{\prime },0\right) =K_d\left( \bar{x}%
^{\prime \prime },\beta |\bar{x}^{\prime },0\right) K_d\left( \bar{y}%
^{\prime \prime },\beta |\bar{y}^{\prime },0\right) K_d\left( \bar{z}%
^{\prime \prime },\beta |\bar{z}^{\prime },0\right) ,  \label{eqKDseparate}
\end{equation}
where the column vector $\bar{x}$ contains the $x$-components of the
particles, i.e. $\bar{x}^T=\left( x_1,\cdots ,x_N\right) $ and similarly for 
$\bar{y}$ and $\bar{z}.$ Knowing the propagator $K\left( x^{\prime \prime
},\beta |x^{\prime },0\right) _\varpi $ of a single harmonic oscillator with
frequency $\varpi $%
\begin{equation}
K\left( x_\beta ,\beta |x_0,0\right) _\varpi =\sqrt{\frac \varpi {2\pi \sinh
\varpi \beta }}\exp \left\{ -\frac \varpi 2\frac{\left( x_\beta
^2+x_0^2\right) \cosh \varpi \beta -2x_\beta x_0}{\sinh \varpi \beta }%
\right\} ,
\end{equation}
one finds for the 1-dimensional propagator $K_d$ of the $N$ distinguishable
oscillators in the interacting system: 
\begin{equation}
K_d\left( \bar{x}^{\prime \prime },\beta |\bar{x}^{\prime },0\right) =\frac{%
K\left( \sqrt{N}X^{\prime \prime },\beta |\sqrt{N}X^{\prime },0\right)
_\Omega }{K\left( \sqrt{N}X^{\prime \prime },\beta |\sqrt{N}X^{\prime
},0\right) _w}\prod_{j=1}^NK\left( x_j^{\prime \prime },\beta |x_j^{\prime
},0\right) _w,  \label{eqKdexplicit}
\end{equation}
where the factor $\sqrt{N}$ in $\sqrt{N}X^{\prime \prime }$ accounts for the
mass $N$ (in atomic units) of the center. The denominator in (\ref
{eqKdexplicit}) compensates for the fact that $\left( N-1\right) $ instead
of $N$ degrees of freedom of frequency $w$ are available. The 3-dimensional
propagator $K_D$ (\ref{eqKDdef}) for $N$ distinguishable oscillators of the
interacting system is, according to (\ref{eqKDseparate}) and (\ref
{eqKdexplicit}), given by 
\begin{equation}
K_D\left( {\bf \bar{r}}^{\prime \prime },\beta |{\bf \bar{r}}^{\prime
},0\right) =\frac{K\left( \sqrt{N}{\bf R}^{\prime \prime },\beta |\sqrt{N}%
{\bf R}^{\prime },0\right) _\Omega }{K\left( \sqrt{N}{\bf R}^{\prime \prime
},\beta |\sqrt{N}{\bf R}^{\prime },0\right) _w}\prod_{j=1}^NK\left( {\bf r}%
_j^{\prime \prime },\beta |{\bf r}_j^{\prime },0\right) _w,
\label{eqKDproduct}
\end{equation}
\begin{equation}
K\left( {\bf r}_j^{\prime \prime },\beta |{\bf r}_j^{\prime },0\right)
_w=K\left( x_j^{\prime \prime },\beta |x_j^{\prime },0\right) _wK\left(
y_j^{\prime \prime },\beta |y_j^{\prime },0\right) _wK\left( z_j^{\prime
\prime },\beta |z_j^{\prime },0\right) _w
\end{equation}
where ${\bf \bar{r}}$ denotes a point in the configuration space $R^{3N},$
i.e. ${\bf \bar{r}}^T\equiv \left( 
\begin{array}{lll}
\left( x_1,y_1,z_1\right) , & \cdots , & \left( x_N,y_N,z_N\right)
\end{array}
\right) $. The symmetrized density matrix $K_I$ for 3D identical particles
(indicated by the subscript $I)$ can be obtained by using the following
projection, with $P$ denoting the permutation matrix: 
\begin{equation}
K_I\left( {\bf \bar{r}}^{\prime \prime },\beta |{\bf \bar{r}}^{\prime
},0\right) =\frac 1{N!}\sum_p\xi ^pK_D\left( P{\bf \bar{r}}^{\prime \prime
},\beta |{\bf \bar{r}}^{\prime },0\right) ,
\end{equation}
where $\xi =+1$ for bosons and $\xi =-1$ for fermions. It should be
emphasized that $P$ acts on the particle indices, not on the components of $%
{\bf r}$ separately. The partition function is then readily obtained by
integrating over the configuration space 
\begin{equation}
Z_I=\int d{\bf \bar{r}}K_I\left( {\bf \bar{r}},\beta |{\bf \bar{r}},0\right)
=\int d{\bf \bar{r}}\frac 1{N!}\sum_p\xi ^pK_D\left( P{\bf \bar{r}},\beta |%
{\bf \bar{r}},0\right) .
\end{equation}

The remaining part of this section will be devoted to the explicit
evaluation of this integral for the partition function. The integration
proceeds in 3 stages: the first stage deals with the center-of-mass
treatment, the second one concerns the cyclic decomposition, and in the
third step the summation over the cycles will be performed.

\subsection{The center of mass}

The center-of-mass coordinate ${\bf R}$ does not only depend on the
coordinates of all the particles, but it also has its own propagator.
Therefore substituting ${\bf R}$ by its expression in terms of the particle
positions and then performing the integration seems not to be the most
adequate way to deal with the integration over the configuration space.
Instead, the following identity is used for the formal treatment of ${\bf R}$
as an independent coordinate, at the expense of additional integrations: 
\begin{equation}
\int d{\bf \bar{r}}f\left( {\bf \bar{r},}\frac 1N\sum_{j=1}^N{\bf r}%
_j\right) =\int d{\bf R}\int d{\bf \bar{r}}f\left( {\bf \bar{r},R}\right)
\delta \left( {\bf R}-\frac 1N\sum_{j=1}^N{\bf r}_j\right) .
\end{equation}
Fourier transformation of the $\delta $-function then leads to 
\begin{equation}
\int d{\bf \bar{r}}f\left( {\bf \bar{r},}\frac 1N\sum_{j=1}^N{\bf r}%
_j\right) =\int d{\bf R}\int \frac{d{\bf k}}{\left( 2\pi \right) ^3}e^{i{\bf %
k}\cdot {\bf R}}\int d{\bf \bar{r}}f\left( {\bf \bar{r},R}\right) e^{-i{\bf 
\bar{k}\cdot \bar{r}}},
\end{equation}
where ${\bf \bar{k}}^T=\frac kN\left( 
\begin{array}{lll}
\left( 1,1,1\right) , & \cdots , & \left( 1,1,1\right)
\end{array}
\right) $ is a $3N$ dimensional row vector. Applying this transformation to
the partition function $Z_I$ and rearranging the factors one obtains 
\begin{equation}
Z_I=\int d{\bf R}\int \frac{d{\bf k}}{\left( 2\pi \right) ^3}e^{i{\bf k}%
\cdot {\bf R}}\frac{K\left( \sqrt{N}{\bf R},\beta |\sqrt{N}{\bf R},0\right)
_\Omega }{K\left( \sqrt{N}{\bf R},\beta |\sqrt{N}{\bf R},0\right) _w}\int d%
{\bf \bar{r}}\frac 1{N!}\sum_p\xi ^p\prod_{j=1}^NK\left( \left( P{\bf r}%
\right) _j,\beta |{\bf r}_j,0\right) _we^{-i\vec{k}{\bf \cdot }\vec{r}_j/N}.
\end{equation}

This transformation makes ${\bf R}$ independent of the particle positions
relative to the center of mass. The real dependence on the relative
positions is reintroduced by the Fourier transform. It should be noted that
the explicit dependence of the propagator (\ref{eqKDproduct}) on ${\bf R,}$
and the presence of the factor $e^{-i\vec{k}{\bf \cdot }\vec{r}_j/N},$ are
consequences of the two-body interactions.

The next step is to rewrite the sum over the permutations as a sum over all
possible cycles. This will be done in the next subsection. An excellent
example of such a decomposition into cycles has been given by Feynman \cite
{FeynStat} for a system of non-interacting particles in a box.

\subsection{Cyclic decomposition}

A permutation can be broken up into cycles. Suppose that a particular
permutation contains $M_\ell $ cycles of length$~\ell .$ The positive
integers $M_\ell $ and $\ell $ then have to satisfy the constraint 
\begin{equation}
\sum_\ell \ell M_\ell =N.  \label{eqMlsum}
\end{equation}
Furthermore, the number $M\left( M_1,\cdots M_N\right) $ of cyclic
decompositions with $M_1$ cycles of length $1,$ $\cdots ,$ $M_\ell $ cycles
of length $\ell ,$ $\cdots $ is known to be 
\begin{equation}
M\left( M_1,\cdots M_N\right) =\frac{N!}{\prod_\ell M_\ell !\ell ^{M_\ell }}.
\end{equation}

A cycle of length $\ell $ will be obtained from $\left( \ell -1\right) $
permutations. Therefore the sign factor $\xi ^p$ can be decomposed as 
\begin{equation}
\xi ^p=\prod_\ell \xi ^{\left( \ell -1\right) M_\ell }.
\end{equation}
Combining these result originating from the permutation symmetry one obtains 
\begin{equation}
Z_I=\int d{\bf R}\int \frac{d{\bf k}}{\left( 2\pi \right) ^3}e^{i{\bf k}%
\cdot {\bf R}}\frac{K\left( \sqrt{N}{\bf R},\beta |\sqrt{N}{\bf R},0\right)
_\Omega }{K\left( \sqrt{N}{\bf R},\beta |\sqrt{N}{\bf R},0\right) _w}%
\sum_{M_1\cdots M_N}\prod_\ell \frac{\xi ^{\left( \ell -1\right) M_\ell }}{%
M_\ell !\ell ^{M_\ell }}\left( {\cal K}_\ell \left( {\bf k}\right) \right)
^{M_\ell },
\end{equation}
\begin{equation}
{\cal K}_\ell \left( {\bf k}\right) =\int d{\bf r}_{\ell +1}\int d{\bf r}%
_\ell \cdots \int d{\bf r}_1\delta \left( {\bf r}_{\ell +1}-{\bf r}_1\right)
\prod_{j=1}^NK\left( {\bf r}_{j+1},\beta |{\bf r}_j,0\right) _we^{-i{\bf %
k\cdot r}_j/N}.
\end{equation}
The $\delta $-function expresses that the decomposition is cyclic. It is
obvious that 
\begin{equation}
{\cal K}_\ell \left( {\bf k}\right) ={\cal K}_\ell ^{\left( 1D\right)
}\left( k_x\right) {\cal K}_\ell ^{\left( 1D\right) }\left( k_y\right) {\cal %
K}_\ell ^{\left( 1D\right) }\left( k_z\right) ,
\end{equation}
which allows to analyze ${\cal K}_\ell \left( {\bf k}\right) $ from its
1-dimensional constituents: 
\begin{equation}
{\cal K}_\ell ^{\left( 1D\right) }\left( k_x\right) =\int dx_{\ell +1}\int
dx_\ell \cdots \int dx_1\delta \left( x_{\ell +1}-x_1\right)
\prod_{j=1}^NK\left( x_{j+1},\beta |x_j,0\right) _we^{-ik_xx_j/N}.
\end{equation}

Using the semigroup property \cite{Roep} of the harmonic oscillator
propagator $K\left( x_{j+1},\beta |x_j,0\right) _w,$ all integrations but
one can be performed 
\begin{equation}
{\cal K}_\ell ^{\left( 1D\right) }\left( k_x\right) =\int dxK\left( x,\ell
\beta |x,0\right) _we^{-\int_0^{\ell \beta }d\tau f_x\left( \tau \right)
x\left( \tau \right) }  \label{eqKl1D}
\end{equation}
where 
\begin{equation}
f_x\left( \tau \right) =i\frac{k_x}N\sum_{j=0}^{\ell -1}\delta \left( \tau
-j\beta \right) .  \label{eqdriving}
\end{equation}
The integral (\ref{eqKl1D}) is the propagator $K_{w,f}$ of a driven harmonic
oscillator with the Lagrangian 
\begin{equation}
L_{w,f_x}=\frac 12\dot{x}^2-\frac 12w^2x^2+f_x\left( \tau \right) x.
\end{equation}
studied in \cite{FeynHibbs,FeynOperator}. It should be noted again that
without two-body interactions the driving force (\ref{eqdriving}) is
lacking. Taking over the result from \cite{FeynOperator} and integrating
over the configuration space one obtains 
\begin{eqnarray}
Z_{w,f_x}\left( \beta \right) &=&\int dxK_{w,f_x}\left( x,\beta |x,0\right) 
\nonumber \\
&=&\frac 1{2\sinh \frac 12\beta w}\exp \left( \frac 12\int_0^\beta d\tau
\int_0^\beta d\sigma \frac{f_x\left( \tau \right) f_x\left( \sigma \right) }{%
2w}\frac{\cosh \left( \left( \frac \beta 2-\left| \tau -\sigma \right|
\right) w\right) }{\sinh \frac 12\beta w}\right) .
\end{eqnarray}

After straightforward algebra one obtains for the 1D function ${\cal K}_\ell
^{\left( 1D\right) }\left( k_x\right) $: 
\begin{equation}
{\cal K}_\ell ^{(1D)}\left( k_x\right) =\frac 1{2\sinh \frac 12\ell \beta w}%
\exp \left( -\frac \ell {4N^2}\frac{k_x^2}w\frac{1+e^{-\beta w}}{1-e^{-\beta
w}}\right) ;
\end{equation}
and for its 3D extension: 
\begin{equation}
{\cal K}_\ell \left( \vec{k}\right) =\left( \frac 1{2\sinh \frac 12\ell
\beta w}\right) ^3\exp \left( -\frac \ell {4N^2}\frac{k^2}w\frac{1+e^{-\beta
w}}{1-e^{-\beta w}}\right) .
\end{equation}

Using (\ref{eqMlsum}) one then is left with a sixfold integral for the
partition function 
\begin{equation}
Z_I= 
\begin{array}[t]{l}
\int d{\bf R}\int \frac{d{\bf k}}{\left( 2\pi \right) ^3}e^{i{\bf k}\cdot 
{\bf R}}\frac{K\left( \sqrt{N}{\bf R},\beta |\sqrt{N}{\bf R},0\right)
_\Omega }{K\left( \sqrt{N}{\bf R},\beta |\sqrt{N}{\bf R},0\right) _w}\exp
\left( -\frac 1{4N}\frac{k^2}w\frac{1+e^{-\beta w}}{1-e^{-\beta w}}\right)
\\ 
\times \sum_{M_1\cdots M_N}\prod_\ell \frac{\xi ^{\left( \ell -1\right)
M_\ell }}{M_\ell !\ell ^{M_\ell }}\left( \frac 1{2\sinh \frac 12\ell \beta w}%
\right) ^{3M_\ell }
\end{array}
.
\end{equation}
Both the integrations over ${\bf k}$ and ${\bf R}$ are Gaussian, leading to
the following series for $Z_I$: 
\begin{equation}
Z_I=\left( \frac{\sinh \frac 12\beta w}{\sinh \frac 12\beta \Omega }\right)
^3{\Bbb Z}_I\left( N\right) ;{\Bbb \quad Z}_I\left( N\right) \equiv
\sum_{M_1\cdots M_N}\prod_\ell \frac{\xi ^{\left( \ell -1\right) M_\ell }}{%
M_\ell !\ell ^{M_\ell }}\left( \frac{e^{-\frac 12\ell \beta w}}{1-e^{-\ell
\beta w}}\right) ^{3M_\ell }.  \label{eqZgeneral}
\end{equation}

Without two-body interactions ($w=\Omega $), ${\Bbb Z}_I\left( N\right) $ is
the partition function of a set of identical oscillators. The partition
function $Z_I$ only differs from it by a center-of-mass correction and the
actual values of $w.$

The remaining summation over the cycles involves the constraint (\ref
{eqMlsum}), which however can be removed by the use of the generating
function technique, which will be considered in the next subsection.

\subsection{The generating function}

Concentrating on the explicit dependence of ${\Bbb Z}_I\left( N\right) $ on $%
N$ (with $w$ considered as a parameter), one can construct the following
generating function 
\begin{equation}
\Xi \left( u\right) =\sum_{N=0}^\infty {\Bbb Z}_I\left( N\right) u^N,
\end{equation}
with ${\Bbb Z}_I\left( 0\right) =1$ by definition. The partition function $%
{\Bbb Z}_I\left( N\right) $ can then be obtained by taking the appropriate
derivatives of $\Xi \left( u\right) $ with respect to $u,$ assuming that the
series for $\Xi \left( u\right) $ is convergent near $u=0$: 
\begin{equation}
{\Bbb Z}_I\left( N\right) =\frac 1{N!}\left. \frac{d^N}{du^N}\Xi \left(
u\right) \right| _{u=0}.
\end{equation}
The summation over the number of cycles with length $\ell $ is now
unrestricted, and can easily be performed: 
\begin{equation}
\Xi _I\left( u\right) =\exp \left( \sum_{\ell =1}^\infty \xi ^{\ell -1}\frac{%
e^{-\frac 32\ell \beta w}u^\ell }{\ell \left( 1-e^{-\ell \beta w}\right) ^3}%
\right) .
\end{equation}
This series can be rewritten into the more familiar form 
\begin{equation}
\Xi _I\left( u\right) =\exp \left( -\xi \sum_{\nu =0}^\infty \frac 12\left(
\nu +1\right) \left( \nu +2\right) \ln \left( 1-\xi ue^{-\beta w\left( \frac %
32+\nu \right) }\right) \right) .  \label{eqKsinomagn}
\end{equation}

It should be noted in view of the remarks at the end of the preceding
section concerning ${\Bbb Z}_I\left( N\right) $, that $\Xi _I\left( u\right) 
$ is also the generating function of a model without two-body interactions.
In that case $w$ equals $\Omega ,$ and $\Xi _I\left( u\right) $ coincides
with the well known \cite{BDLSSCpress,Grossmann,Kirsten} grand canonical
partition function of a set of identical particles in a parabolic well.

\subsection{Recurrence relations for the partition function}

Starting from the expression for $\Xi _I\left( u\right) $ derived in the
previous subsection for the interacting model, a recursion relation can be
obtained for ${\Bbb Z}_I\left( N\right) $. Introducing: 
\begin{equation}
b=e^{-\beta w}
\end{equation}
for brevity in the notations, we observe that 
\begin{equation}
\frac d{du}\Xi _I\left( u\right) =\Xi _I\left( u\right) \sum_{\nu =0}^\infty 
\frac 12\left( \nu +1\right) \left( \nu +2\right) \frac{b^{\frac 32+\nu }}{%
1-\xi ub^{\frac 32+\nu }}
\end{equation}
Considering next ${\Bbb Z}_I\left( N\right) =\frac 1{N!}\frac{d^{N-1}}{%
du^{N-1}}\left. \frac d{du}\Xi \left( u\right) \right| _{u=0},$ the product
rule and an elementary binomial expansion can be used to find 
\begin{equation}
{\Bbb Z}_I\left( N\right) =\frac 1N\sum_{m=0}^{N-1}\xi ^{N-m-1}\left( \frac{%
b^{\frac 12\left( N-m\right) }}{1-b^{N-m}}\right) ^3{\Bbb Z}_I\left(
m\right) .  \label{ZIrecur}
\end{equation}

The corresponding 1-dimensional version of this recurrence relation
(indicated with the subscript $i$ to distinguish it from the 3D case with
capital subscript) becomes 
\begin{equation}
{\Bbb Z}_i\left( N\right) =\frac 1N\sum_{m=0}^{N-1}\xi ^{N-m-1}\frac{b^{%
\frac 12\left( N-m\right) }}{1-b^{N-m}}{\Bbb Z}_i\left( m\right) ,
\end{equation}
leading to the following partition functions in closed form for
one-dimensional bosons and one-dimensional fermions 
\begin{equation}
{\Bbb Z}_b=\frac{b^{\frac 12N}}{\prod_{j=1}^N\left( 1-b^j\right) };\quad 
{\Bbb Z}_f=\frac{b^{\frac 12N^2}}{\prod_{j=1}^N\left( 1-b^j\right) }.
\end{equation}
It is easy to check that these partition functions are the solution of the
recurrence relation for ${\Bbb Z}_i\left( N\right) $ with $\xi =1$ for
bosons and $\xi =-1$ for fermions. However we did not find a systematic
method to obtain analytical solutions of this type of recurrence relations.
E.g. for the 3D case we had to rely on numerical schemes, as will be
discussed below. But at this stage, it is worthwhile first to consider the
presence of an homogeneous magnetic field, as the origin of anisotropy in
our model of $N$ identical oscillators.

\section{$N$ identical oscillators in a magnetic field}

The Lagrangian of $N$ particles in a confining parabolic potential in the
presence of a magnetic field is 
\begin{equation}
L_{\omega _c}=\frac 12\sum_{j=1}^N\left( {\bf \dot{r}}_j-2\omega _cx_j\dot{y}%
_j\right) ^2-\frac 12\Omega ^2\sum_{j=1}^N{\bf r}_j^2,  \label{eqLmagn}
\end{equation}
where $\omega _c$ is the cyclotron frequency. For this model, the
calculations of the preceding section can in essence be repeated. First the
propagator for distinguishable particles is calculated. The next step will
be the projection on the irreducible representation of the permutation
group, and performing the cyclic decomposition. Then the generating function
is introduced to circumvent the constraints on the partition in cycles, and
finally the summation over the cycles is performed.

The fact that the energy spectrum and the wavefunction can be calculated
when harmonic interparticle interactions are included \cite{Johnson}
indicates that the propagator and the partition function for the model in a
magnetic field with two-body interaction can be obtained using our methods.
The calculation technique is very demanding, therefore it seemed appropriate
to illustrate the method at the hand of the simple model (\ref{eqLmagn}).

\subsection{The propagator for distinguishable particles}

For distinguishable particles, the many-particle propagator is a product of
one-particle propagators. The one-particle propagator $K_{\omega _c}^{\left(
1\right) }\left( {\bf r},\beta |{\bf r}^{\prime }\right) $ of this model can
be calculated by stochastic techniques\cite{Simon} or by path integral
techniques\cite{FeynHibbs}. The evaluation is somewhat lengthy but
straightforward, resulting eventually in 
\begin{equation}
K_{\omega _c}^{\left( 1\right) }\left( {\bf r},\beta |{\bf r}^{\prime
}\right) = 
\begin{array}[t]{l}
\sqrt{\frac \Omega {2\pi \sinh \beta \Omega }}\frac s{2\pi \sinh \beta s}%
\exp \left\{ -\frac{\Omega \left( \left( z^2+\left( z^{\prime }\right)
^2\right) \cosh \beta \Omega -2zz^{\prime }\right) }{2\sinh \beta \Omega }%
\right\} \\ 
\times \exp \left\{ -\frac s2\frac{\left( x^2+y^2+(x^{\prime })^2+(y^{\prime
})^2\right) \cosh \beta s-2\left( xx^{\prime }+yy^{\prime }\right) \cosh 
\frac 12\beta \omega }{\sinh \beta s}\right\} \\ 
\times \exp \left\{ -i\left( \frac 12\omega _c\left( xy-x^{\prime }y^{\prime
}\right) -s\frac{\sinh \frac 12\beta \omega }{\sinh \beta s}\left( y^{\prime
}x-yx^{\prime }\right) \right) \right\} ,
\end{array}
\end{equation}
with the eigenfrequency $s$ given by 
\begin{equation}
s=\sqrt{\Omega ^2+\frac 14\omega _c^2}.
\end{equation}
The partition function corresponding to this single-particle propagator is
obtained by integrating over the configuration space: 
\begin{equation}
{\Bbb Z}_{\omega _c}^{\left( 1\right) }\left( \beta \right) =\int d{\bf r}%
K_{\omega _c}^{\left( 1\right) }\left( {\bf r},\beta |{\bf r}\right) =\left[
8\sinh \beta \frac \Omega 2\sinh \beta \left( \frac s2+\frac{\omega _c}4%
\right) \sinh \beta \left( \frac s2-\frac{\omega _c}4\right) \right] ^{-1}
\label{eqZmagn}
\end{equation}
which coincides with the partition function of a 3D harmonic oscillator if $%
\omega _c=0.$

\subsection{The partition function and generating function for identical
particles}

For $N$ identical particles, the propagator becomes 
\begin{equation}
K_{I,\omega _c}\left( {\bf \bar{r}},\beta |{\bf \bar{r}}^{\prime }\right) =%
\frac 1{N!}\sum_p\xi ^p\prod_{j=1}^NK_{\omega _c}^{\left( 1\right) }\left(
\left( P{\bf r}\right) _j,\beta |{\bf r}_j^{\prime }\right) ,
\end{equation}
with the corresponding partition function given by 
\begin{equation}
{\Bbb Z}_{I,\omega _c}\left( N\right) =\int d{\bf \bar{r}}K_{I,\omega
_c}\left( {\bf \bar{r}},\beta |{\bf \bar{r}}\right) .
\end{equation}

Using the cyclic decomposition and the semigroup property as before,
followed by the relaxation of the constraint (\ref{eqMlsum}) on the number
of cycles, one finds for the generating function 
\begin{equation}
\Xi _{\omega _c}\left( u\right) =\exp \left( \sum_{\ell =1}^\infty \frac{\xi
^{\ell -1}}\ell {\Bbb Z}_{\omega _c}^{\left( 1\right) }\left( \ell \beta
\right) u^\ell \right) .  \label{eqKsimagn}
\end{equation}

Because in this model with a magnetic field no two-body interactions were
taken into account, this expression equals the grand-canonical partition
function of the model if $u=e^{\beta \mu }$ is interpreted as the fugacity,
with $\mu $ the chemical potential.

We have thus obtained an expression for the grand-canonical partition
function in the absence of two-body interactions for $N$ identical harmonic
oscillators, both with and without a magnetic field. For a system with
harmonic two-body interactions, repulsive or attractive, expressions for the
canonical partition function are obtained (in the absence of a magnetic
field). The corresponding thermodynamic properties (grand-canonical ground
state occupancy, canonical and grand-canonical specific heat and average
square radius) of the confined system will be calculated in the next section.

\section{Identical particles in 1D}

In the preceding sections we have obtained the grand canonical partition
function of identical harmonic oscillators in a confining parabolic
potential well without a magnetic field [equation (\ref{eqKsinomagn})] and
in the presence of a magnetic field [equation (\ref{eqKsimagn})]. For the
same system the canonical partition function is obtained for an attractive
as well as for a repulsive harmonic two-body interaction [equation (\ref
{eqZgeneral})]. The expressions are given for bosons and for fermions. In
this section we will analyze the thermodynamical properties of the confined
system in the case of motion in 1D. It should be noted that for the
non-interacting case in 1D, we recover the results of Ref. 
\onlinecite{Takahashi}%
. To the best of our knowledge, the interacting case has not been analyzed
up to now for identical particles. For distinguishable particles it has been
studied before \cite{Lieb}.

In 1D, the explicit expressions for $Z$ largely facilitate the analysis: the
free energy $F=-\frac 1\beta \ln Z,$ the internal energy $U=\frac{d\beta F}{%
d\beta }$ and the specific heat $C=\frac{dU}{dT}$ can be readily obtained.
The results are summarized in Table I. The subscript $CM\;$refers to the
center-of-mass contribution.

The center-of-mass correction $C_{CM}$ to the specific heat --which is an
excess specific heat and therefore may be negative-- is shown in Fig. 1. In
Fig. 2 we plot the contribution $\Delta C$ to the specific heat, due to the
internal degrees of freedom. Note that $\Delta C$ is the same for 1D
fermions and bosons; bosonization implies this result \cite{Tomonaga}.

The spatial extension of the cloud of interacting identical particles can be
obtained by applying the Hellman-Feynman theorem to the free energy: 
\begin{equation}
\frac{\partial F}{\partial \left( \Omega ^2\right) }=\frac 1{2Z}\text{Tr}%
\left( e^{-\beta H}\sum_{j=1}^Nx_j^2\right) =\frac N2\left\langle
x^2\right\rangle .
\end{equation}
The difference $\left\langle x^2\right\rangle _f-\left\langle
x^2\right\rangle _b$ between the mean square length of the 1D fermion and
boson system turns out to be: 
\begin{equation}
\left\langle x^2\right\rangle _f-\left\langle x^2\right\rangle _b=\frac{N-1}{%
2w}.
\end{equation}
It is proportional to the number of particles, and inversely proportional to
the frequency $w$ of the internal degrees of freedom.

\section{Bosons in 3D}

The calculation of the thermodynamic properties or the mean square radius of
the 3D cloud of identical particles is substantially complicated by the lack
of an explicit expression in closed form for the partition function.
Numerical methods have therefore to be used. We concentrate this analysis on
the boson case. For fermions a different scheme will have to be developed
and will be published later.

\subsection{In the absence of a magnetic field}

The recurrence relation (\ref{ZIrecur}) is not directly accessible for
numerical computation, as can easily be seen by evaluating the expected
dominant factor $b^{3N/2}/\prod_{j=1}^N\left( 1-b^j\right) ^3$ for bosons.
For a relatively low temperature and a moderate number of particles, say $%
b=0.75$ and $N=1000,$ this factor is as small as $1.0402\times 10^{-182}.$
We therefore isolate this factor using the following scaling for the
partition function 
\begin{equation}
{\Bbb Z}_B\left( N\right) =\sigma _N\frac{b^{\frac 32N}}{\prod_{j=1}^N\left(
1-b^j\right) ^3},
\end{equation}
and rewrite the recurrence relation (\ref{ZIrecur}) in terms of the activity 
$\rho _N$ (see e.g. \cite{Lieb}) defined as 
\begin{equation}
\sigma _N=\rho _N\sigma _{N-1}\Longrightarrow \sigma _N=\sigma
_0\prod_{j=0}^N\rho _j
\end{equation}
where $\sigma _0=\rho _0\equiv 1$ have been introduced for convenience. The
recurrence relation (\ref{ZIrecur}) then becomes after some manipulations 
\begin{equation}
\rho _N=\frac 1N\left( \frac{1-b^N}{1-b}\right) ^3\left(
1+\sum_{m=0}^{N-2}\left( \frac{1-b}{1-b^{N-m}}\right) ^3\prod_{j=m+1}^{N-1}%
\frac{\left( 1-b^j\right) ^3}{\rho _j}\right) .  \label{eqrhorecur}
\end{equation}
The corresponding recurrence relation for the internal energy of the
internal degrees of freedom becomes 
\begin{equation}
\frac{{\Bbb U}_B\left( N\right) }{\hbar w}=\frac 1N\frac 1{\rho _N}\left( 
\frac{1-b^N}{1-b}\right) ^3\left( 
\begin{array}{l}
\frac 32\frac{1+b}{1-b}+\frac{{\Bbb U}_B\left( N-1\right) }{\hbar w} \\ 
+\sum_{m=0}^{N-2}\left( \left( \frac 32\left( N-m\right) \frac{1+b^{N-m}}{%
1-b^{N-m}}+\frac{{\Bbb U}_B\left( m\right) }{\hbar w}\right) \left( \frac{1-b%
}{1-b^{N-m}}\right) ^3\prod_{j=m+1}^{N-1}\frac{\left( 1-b^j\right) ^3}{\rho
_j}\right)
\end{array}
\right) .  \label{eqUrecur}
\end{equation}
The temperature scale used to express $b=e^{-w/kT}$ is 
\begin{equation}
t=\left( \frac N{\zeta \left( 3\right) }\right) ^{-1/3}\frac{kT}w\equiv 
\frac T{T_c}\text{ where }\zeta \left( 3\right) =1.\,2021.
\end{equation}

The routine used to calculate the values $\rho _1\cdots \rho _N$ and ${\Bbb U%
}_B\left( 1\right) \cdots {\Bbb U}_B\left( N\right) $ from these recurrence
relations (\ref{eqrhorecur}) and (\ref{eqUrecur}) is displayed in Table II.

The values of $\rho _1\cdots \rho _N$ for $N=100$ are shown in Fig. 3 for
the three temperatures $T/T_c=0.5,$ 1 and 2. For $T=2T_c$ it turns out that $%
\rho _j$ is about 8 times larger than $\rho _{j-1}$ for $j$ approaching $N,$
which makes it extremely difficult to deal numerically with the values of
the partition function rather than with the proportionality factors.

Once $\rho _N$ is known, the internal energy ${\Bbb U}_B\left( N\right) $ is
readily obtained from (\ref{eqUrecur}). The results for $N=10,$ 100 and 1000
are shown in Fig. 4. The specific heat in the canonical ensemble can also be
calculated from these parameters. This is shown in Fig. 5, clearly
illustrating the effect of condensation for a finite number of particles.
Anticipating the next subsection, it should be noted that the specific heat
in the canonical ensemble for $N$ particles and without repulsive two-body
interactions is identical to the specific heat in the grand canonical
ensemble with the average number of particles given by the same value of $N.$
The reason why we compare both ensembles without repulsive interactions is
the requirement that in the grand canonical ensemble the system should be
stable for any large number of particles, which is not the case in the
Gaussian model with repulsion, because the confining potential can only
accommodate a finite number of particles as a consequence of (\ref{eqeigenw}%
). Another interesting consequence of the repulsive interactions follows
from the dependence of the condensation temperature $T_c$ on the number of
particles, which obeys the following scaling law taking (\ref{eqeigenw})
into account: 
\begin{equation}
T_c=\frac{\sqrt{\Omega ^2-N\omega ^2}}k\left( \frac N{\zeta \left( 3\right) }%
\right) ^{1/3}\Longrightarrow \frac{kT_c}\Omega \left( \frac{\omega ^2}{%
\Omega ^2}\zeta \left( 3\right) \right) ^{1/3}=\sqrt{1-\frac{N\omega ^2}{%
\Omega ^2}}\left( \frac{N\omega ^2}{\Omega ^2}\right) ^{1/3}
\end{equation}
For the case of attractive interactions and no confinement potential, $T_c$
is proportional to $N^{4/3}.$ The condensation temperature for both cases is
plotted in Fig. 6. For the Gaussian model, the case of harmonic attraction
does not pose any problem because in this model there is no sign of an
``extra'' collapse due to the nature of the interaction. The only
consequence seems to be that the condensation occurs at a much higher
temperature than would be the case without two-body interactions or with a
repulsive harmonic interaction.

\subsection{In the presence of a magnetic field}

The thermodynamic properties in the grand canonical ensemble of an ideal
Bose gas in a parabolic well and in the presence of a magnetic field can be
obtained directly from (\ref{eqZmagn}) and (\ref{eqKsimagn}). Substituting $%
u $ by the fugacity $e^{\beta \mu },$ the Gibbs free energy $G_{\omega _c}$
for the boson case ($\xi =+1$) becomes: 
\begin{equation}
G_{\omega _c}=-\frac 1\beta \sum_{\ell =1}^\infty \frac 1\ell \frac{e^{\ell
\beta \left( \mu -\frac 12\Omega -s\right) }}{\left( 1-e^{-\ell \beta \Omega
}\right) \left( 1-e^{-\ell \beta \left( s+\frac 12\omega _c\right) }\right)
\left( 1-e^{-\ell \beta \left( s-\frac 12\omega _c\right) }\right) }.
\end{equation}
After a power series expansion of the denominators, the summation over the
cycle lengths $\ell $ can be performed: 
\begin{equation}
G_{\omega _c}=\frac 1\beta \sum_{j,k,l=0}^\infty \ln \left( 1-e^{\beta \mu
}e^{-\beta \Omega \left( \frac 12+j\right) }e^{-\beta s\left( 1+k+l\right)
}e^{-\frac 12\beta \omega _c\left( k-l\right) }\right) ,
\end{equation}
as expected from a Bose-Einstein distribution with energy levels 
\begin{equation}
\epsilon _{j,k,l}=\frac 12\Omega +s+j\Omega +k\left( s+\frac 12\omega
_c\right) +l\left( s-\frac 12\omega _c\right) .
\end{equation}

The average number of particles $N=-\frac{\partial G_{\omega _c}}{\partial
\mu }$ is given by 
\begin{equation}
N=\sum_{j,k,l=0}^\infty n_{j,k,l};\quad n_{j,k,l}=\frac{e^{\beta \mu
}e^{-\beta \Omega \left( \frac 12+j\right) }e^{-\beta s\left( 1+k+l\right)
}e^{-\frac 12\beta \omega _c\left( k-l\right) }}{1-e^{\beta \mu }e^{-\beta
\Omega \left( \frac 12+j\right) }e^{-\beta s\left( 1+k+l\right) }e^{-\frac 12%
\beta \omega _c\left( k-l\right) }},
\end{equation}
and the fugacity can be eliminated in favor of the ground state occupancy $%
n_0$: 
\begin{equation}
n_0=\frac{e^{\beta \mu }e^{-\frac 12\beta \Omega }e^{-\beta s}}{1-e^{\beta
\mu }e^{-\frac 12\beta \Omega }e^{-\beta s}}\Longrightarrow e^{\beta \mu
}=\alpha e^{\beta \left( \frac 12\Omega +s\right) }\text{ with }\alpha
\equiv \frac{n_0}{n_0+1}
\end{equation}
Restoring the cyclic summation for reasons of numerical convergence, one
obtains for the average number of particles 
\begin{equation}
N=\sum_{\ell =1}^\infty \frac{\alpha ^\ell }{D_\ell };\quad D_\ell \equiv
\left( 1-e^{-\ell \beta \Omega }\right) \left( 1-e^{-\ell \beta \left( s+%
\frac 12\omega _c\right) }\right) \left( 1-e^{-\ell \beta \left( s-\frac 12%
\omega _c\right) }\right) ,
\end{equation}
which can be cast in the form of the following numerically tractable series 
\begin{equation}
N=\frac \alpha {1-\alpha }+\sum_{\ell =1}^\infty \alpha ^\ell \left( \frac 1{%
D_\ell }-1\right) .
\end{equation}
For given $N,$ standard numerical techniques can be used to determine $%
\alpha \in \left[ 0,1\right] ,$ and hence the ground state occupancy $n_0,$
which is shown in Fig. 7 for $\omega _c=0$ and in Fig. 8 for $\omega
_c/\Omega =5$ as a function of $T/T_c$ for several values of $N.$ It turns
out that the parabolic well is more important than the anisotropy due to the
presence of the magnetic field. The anisotropy only moderately influences
the temperature dependence of the ground state occupancy.

The internal energy $U_{\omega _c}=\frac{\partial \left( \beta G_{\omega
_c}\right) }{\partial \beta }-\mu N$ becomes 
\begin{equation}
U_{\omega _c}=\sum_{\ell =1}^\infty \frac{\alpha ^\ell }{D_\ell }\left(
\Omega \frac{e^{-\ell \beta \Omega }}{1-e^{-\ell \beta \Omega }}+\left( s+%
\frac 12\omega _c\right) \frac{e^{-\ell \beta \left( s+\frac 12\omega
_c\right) }}{1-e^{-\ell \beta \left( s+\frac 12\omega _c\right) }}+\left( s-%
\frac 12\omega _c\right) \frac{e^{-\ell \beta \left( s-\frac 12\omega
_c\right) }}{1-e^{-\ell \beta \left( s-\frac 12\omega _c\right) }}\right)
\end{equation}
and its numerical evaluation once $\alpha $ is determined presents no
numerical difficulties. The resulting specific heat is shown in Fig. 9 for $%
\omega _c/\Omega =5.$ Comparison with Fig. 5 reveals that the anisotropy due
to the magnetic field essentially broadens the peak in the specific heat
near the condensation temperature, but does not substantially alter the
structure.

\section{Discussion and conclusion}

In this paper we applied the method of symmetrical density matrices,
developed by Feynman for a system of non-interacting particles in a box, to
a system of harmonically interacting particles in a confining parabolic
potential. The interaction could be taken into account, due to the Gaussian
nature of the propagators, allowing integration over the configuration
space. The symmetrization resulting from the projection of the propagators
for distinguishable particles on the appropriate representation of the
permutation group gives rise to a series which could be summed using the
generating function technique. Without these generating functions the
calculation has to be restricted to a limited number of particles \cite
{Haase,Hausler,Hammermesh}. Using them not only the grand canonical
partition function $\Xi \left( u\right) $ could be obtained in the parameter
range of the model where $\Xi \left( u\right) $ is well defined, but also
the canonical partition functions $Z\left( N\right) $ for a given number $N$
of identical particles could be obtained as a recursion of partition
functions of a smaller number of particles for the interacting system. The
recurrence relation for the activity $\rho _N,$ i.e. the proportionality
factor between $Z\left( N\right) $ and $Z\left( N-1\right) ,$ allows for an
accurate numerical treatment of the thermodynamical quantities of the model,
such as the internal energy and the specific heat.

Also the thermodynamic properties of the same model in the presence of a
homogeneous magnetic field could be investigated along these lines. A
detailed analysis of the ideal gas in a confining parabolic potential with
anisotropy due to a magnetic field was presented; special attention was
payed to the relation between the number of condensed atoms, the magnetic
field, the strength of the confinement potential and the total number of
particles. The relationship between the parameters of our model and the
characteristics of atomic traps \cite{BEC1,BEC2,BEC3} lies beyond the scope
of the present paper.

It should be mentioned that in the absence of two-body interactions, the
generating function can be identified as the grand canonical partition
function. In that case the specific heat as obtained from the grand
canonical ensemble for an average number of $\left\langle N\right\rangle $
particles calculated using the chemical potential equals the specific heat
obtained from the canonical ensemble with the number of particles $N$ given
by $\left\langle N\right\rangle .$ For a more elaborate discussion we refer
to \cite{BDLSSCpress}.

It should also be mentioned that in this the model it is assumed that the
spin degrees of freedom are fixed. This simplifying assumption is imposed by
the symmetrization method, which becomes more involved if the spin degrees
of freedom depend on the configuration of the particles.

In summary, the partition function of a general Gaussian model for bosons
and fermions, with or without a magnetic field, has been calculated
analytically, and the thermodynamical properties of this model have been
studied. Particular attention has been given to the Bose-Einstein
condensation in the presence of two-body interactions, repulsive and
attractive, and in the presence of a magnetic field.

\acknowledgments
Part of this work is performed in the framework of the NFWO projects No.
2.0093.91, 2.0110.91, G. 0287.95 and WO.073.94N (Wetenschappelijke
Onderzoeksgemeenschap, Scientific Research Community of the NFWO on
``Low-Dimensional Systems''), and in the framework of the European Community
Program Human Capital and Mobility through contracts no. CHRX-CT93-0337 and
CHRX-CT93-0124. One of the authors (F.B.) acknowledges the National Fund for
Scientific Research for financial support.

\newpage

\begin{center}
{\bf Tables}
\end{center}

\begin{table}[tbp] \centering%
\begin{tabular}{|l|l|l|l|l|}
\hline
& Center-of-mass contribution & $i=f$ (fermions) & $i=b$ (bosons) &  \\ 
\hline
$Z=Z_{CM}{\Bbb Z}_i$ & $\frac{\sinh w\beta /2}{\sinh \Omega \beta /2}$ & $%
\frac{e^{-\beta wN^2/2}}{\prod_{j=1}^N\left( 1-e^{-\beta wj}\right) }$ & $%
\frac{e^{-\beta wN/2}}{\prod_{j=1}^N\left( 1-e^{-\beta wj}\right) }$ &  \\ 
\hline
$F=F_{CM}+{\Bbb F}_i$ & $\frac{\Omega -w}2+\frac 1\beta \ln \frac{%
1-e^{-\Omega \beta }}{1-e^{-w\beta }}$ & $\frac 12N^2w+\Delta F$ & $\frac 12%
Nw+\Delta F$ & $\Delta F=\frac 1\beta \sum_{j=1}^N\ln \left( 1-e^{-\beta
wj}\right) $ \\ \hline
$U=U_{CM}+{\Bbb U}_i$ & $\frac{\Omega -w}2+\frac{\Omega e^{-\Omega \beta }}{%
1-e^{-\Omega \beta }}-\frac{we^{-w\beta }}{1-e^{-w\beta }}$ & $\frac 12%
N^2w+\Delta U$ & $\frac 12Nw+\Delta U$ & $\Delta U=w\sum_{j=1}^N\frac{%
je^{-\beta wj}}{1-e^{-\beta wj}}$ \\ \hline
$C=C_{CM}+{\Bbb C}_i$ & $k\beta ^2\left( \frac{\Omega ^2e^{-\Omega \beta }}{%
\left( 1-e^{-\Omega \beta }\right) ^2}-\frac{w^2e^{-w\beta }}{\left(
1-e^{-w\beta }\right) ^2}\right) $ & $\Delta C$ & $\Delta C$ & $\Delta
C=k\beta ^2w^2\sum_{j=1}^N\frac{j^2e^{-\beta wj}}{\left( 1-e^{-\beta
wj}\right) ^2}$ \\ \hline
\end{tabular}
\caption{Partition function, free energy, internal energy and specific heat of 1D identical particles in a parabolic confining potential with harmonic two-body interactions.\label{table1}}%
\end{table}%

\begin{table}[tbp] \centering%
\hrulefill%
\begin{verbatim}
      subroutine recur(NMAX,b,RHO,U)
      dimension RHO(0:NMAX),U(0:NMAX)
      RHO(1)=1
      U(1)=1.5*(1+b)/(1-b)
      if (NMAX.le.1) return
      do N=1,NMAX
         RHO(N)=0
         do   m=0,N-2
            RHO(N)=(1-b**(m+1))**3/RHO(m+1)*
     *       ( RHO(N)+( (1-b)/(1-b**(N-m)) )**3 )
         enddo
         RHO(N)=(1+RHO(N))/N *((1-b**N)/(1-b))**3
         U(N)=0
         do   m=0,N-2
            DU=1.5*(N-m)*(1+b**(N-m))/(1-b**(N-m)) + U(m)
            U(N)=(1-b**(m+1))**3/RHO(m+1)*
     *       ( U(N)+DU*( (1-b)/(1-b**(N-m)) )**3 )
         enddo
         U(N)=1./N/RHO(N)*((1-b**N)/(1-b))**3*
     *       (U(N)+1.5*(1+b)/(1-b)+U(N-1))
      enddo
      end
\end{verbatim}

\hrulefill%
\caption{Subroutine to calculate the values $\rho _1\cdots \rho _N$ and ${\Bbb{U}}_B\left( 1\right) \cdots {\Bbb{U}}_B\left( N\right) $ from the recurrence relations (\ref{eqrhorecur}) and (\ref{eqUrecur}).\label{table2}}%
\end{table}%

\begin{center}
{\bf Figure captions}
\end{center}

\begin{description}
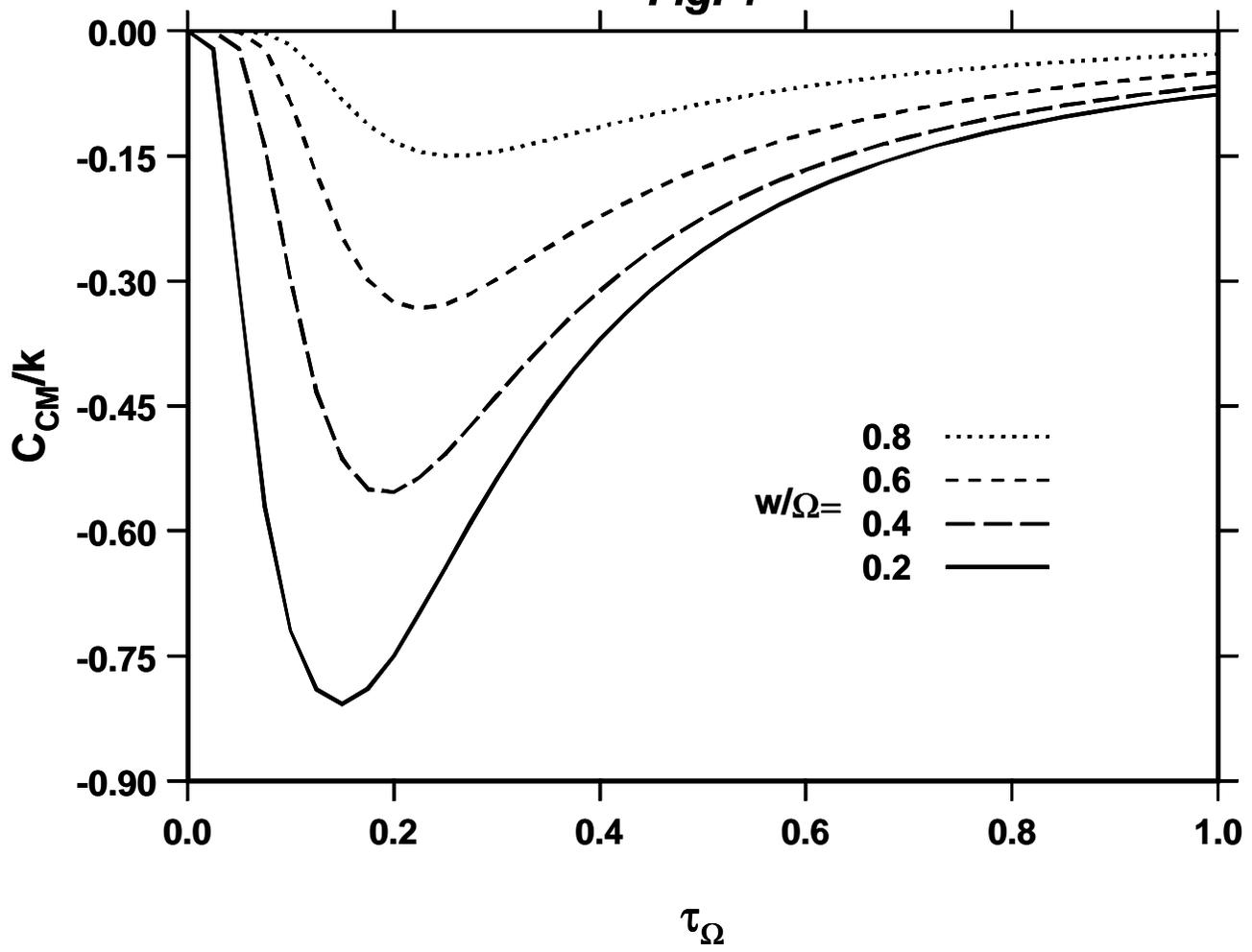
\item[Fig. 1:]  Center-of-mass correction $C_{CM}$ to the specific heat of
1D fermions and bosons (see table I) in units of the Boltzmann constant $k$
as a function of the reduced temperature $\tau _\Omega =kT/\Omega ,$ for
various ratios of $w/\Omega .$

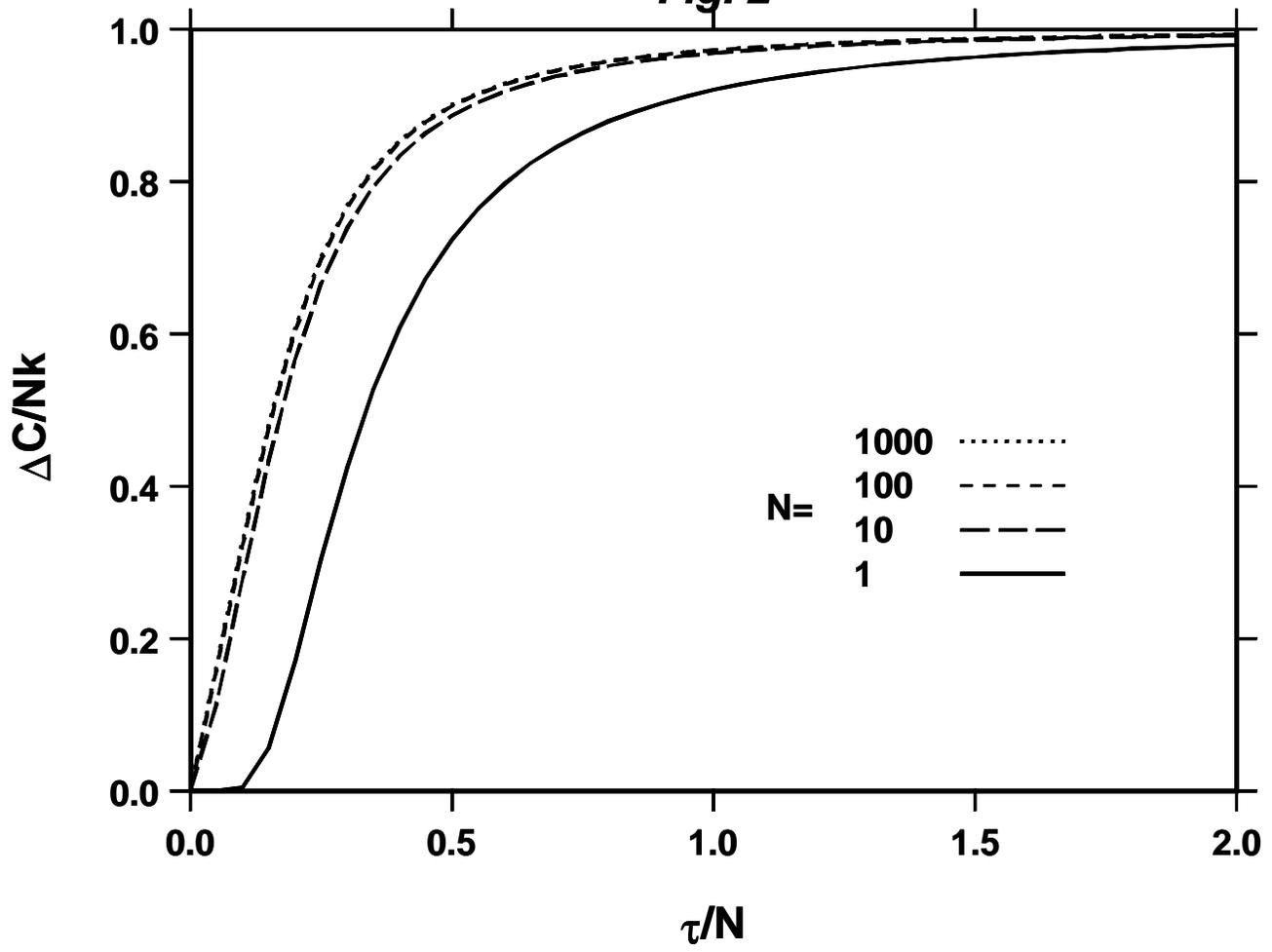
\item[Fig. 2:]  Specific heat contribution $\Delta C/Nk$ per particle for 1D
fermions and bosons (see table I) in units of the Boltzmann constant $k$ as
a function of $\tau /N$ with $\tau $ the reduced temperature $\tau =kT/w,$
for several values of $N$.

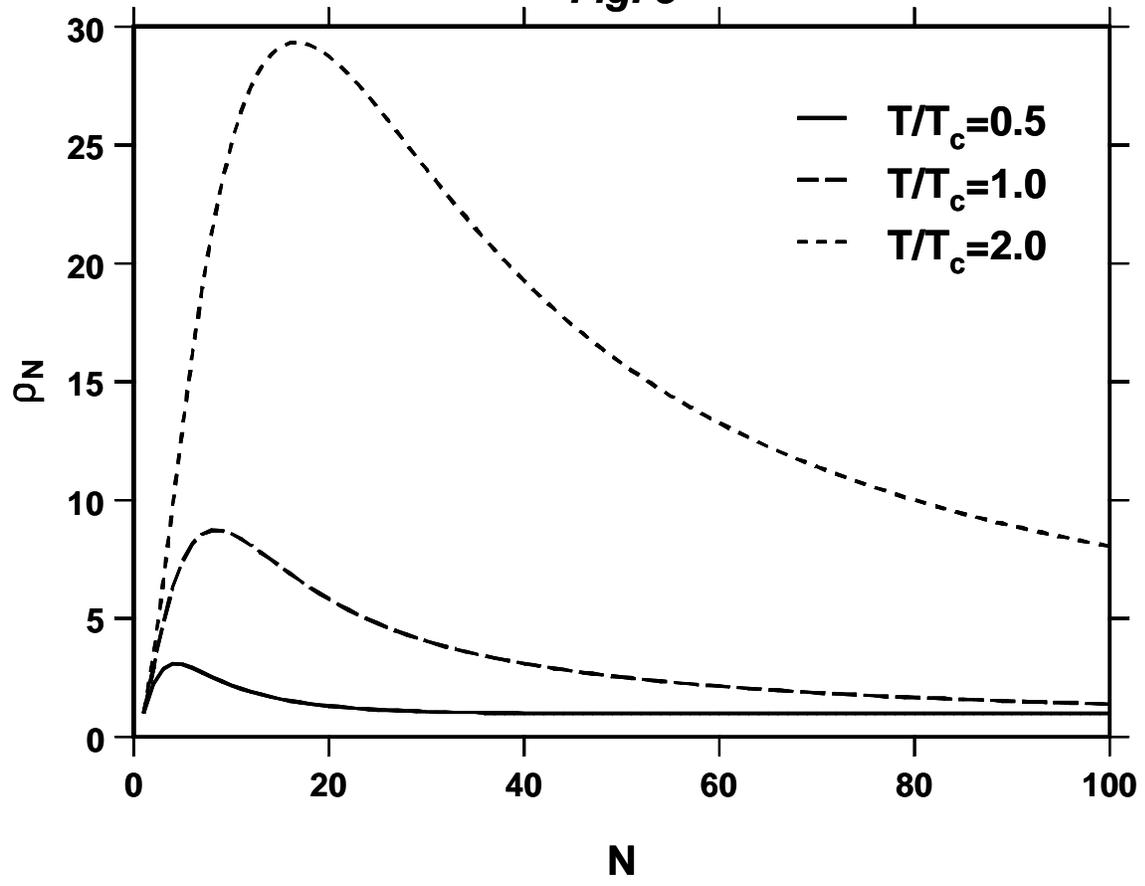
\item[Fig. 3:]  Values of $\rho _1\cdots \rho _N$ for $N=100$ from the
recurrence relation (\ref{eqrhorecur}) for the three temperatures $%
T/T_c=0.5, $ $1$ and $2$.

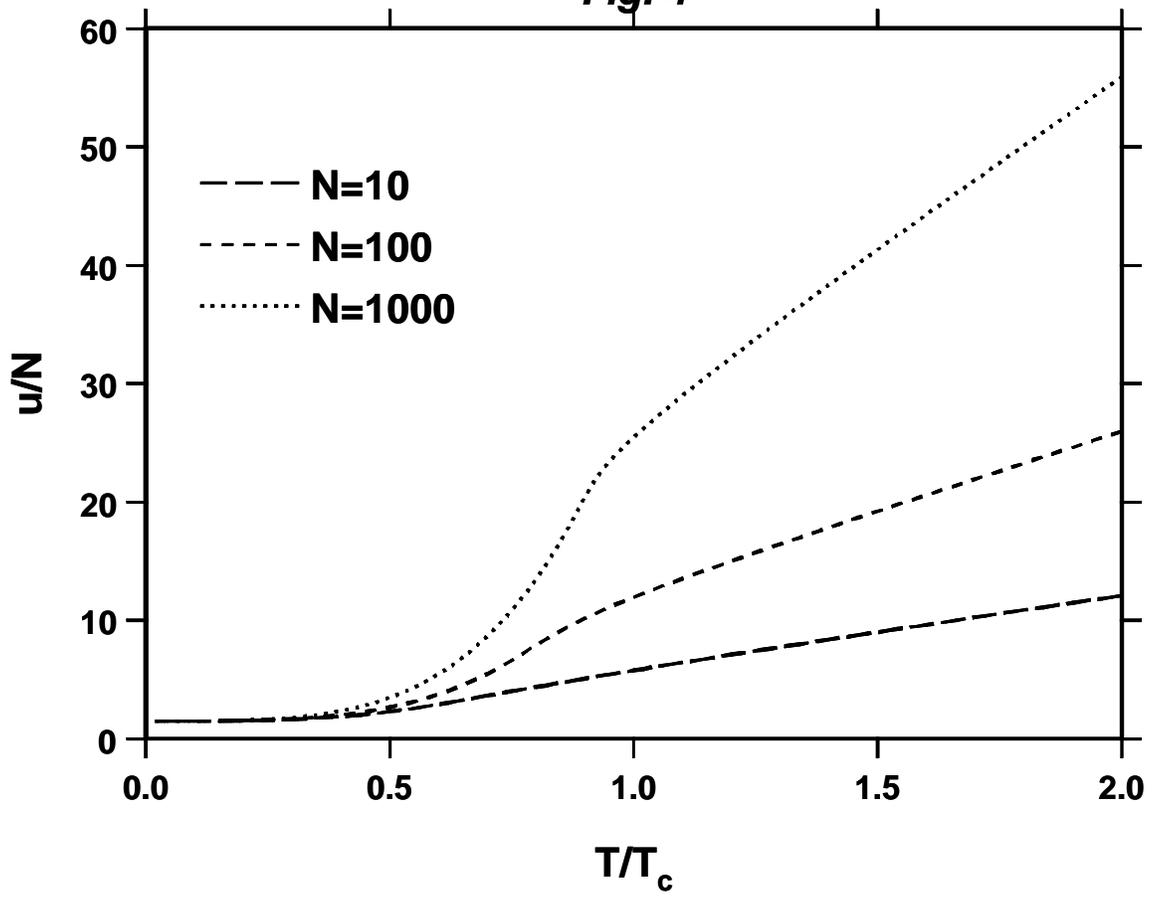
\item[Fig. 4:]  Internal energy $u/N\equiv {\Bbb U}_B\left( N\right) /N\hbar
w$ for $N=10,100,1000$ as a function of $T/T_c.$

\item[Fig. 5:]  Specific heat ${\bf C}_B/Nk$ per particle in units of the
Boltzmann constant $k$ for $N=10,100,1000$ as a function of $T/T_c.$

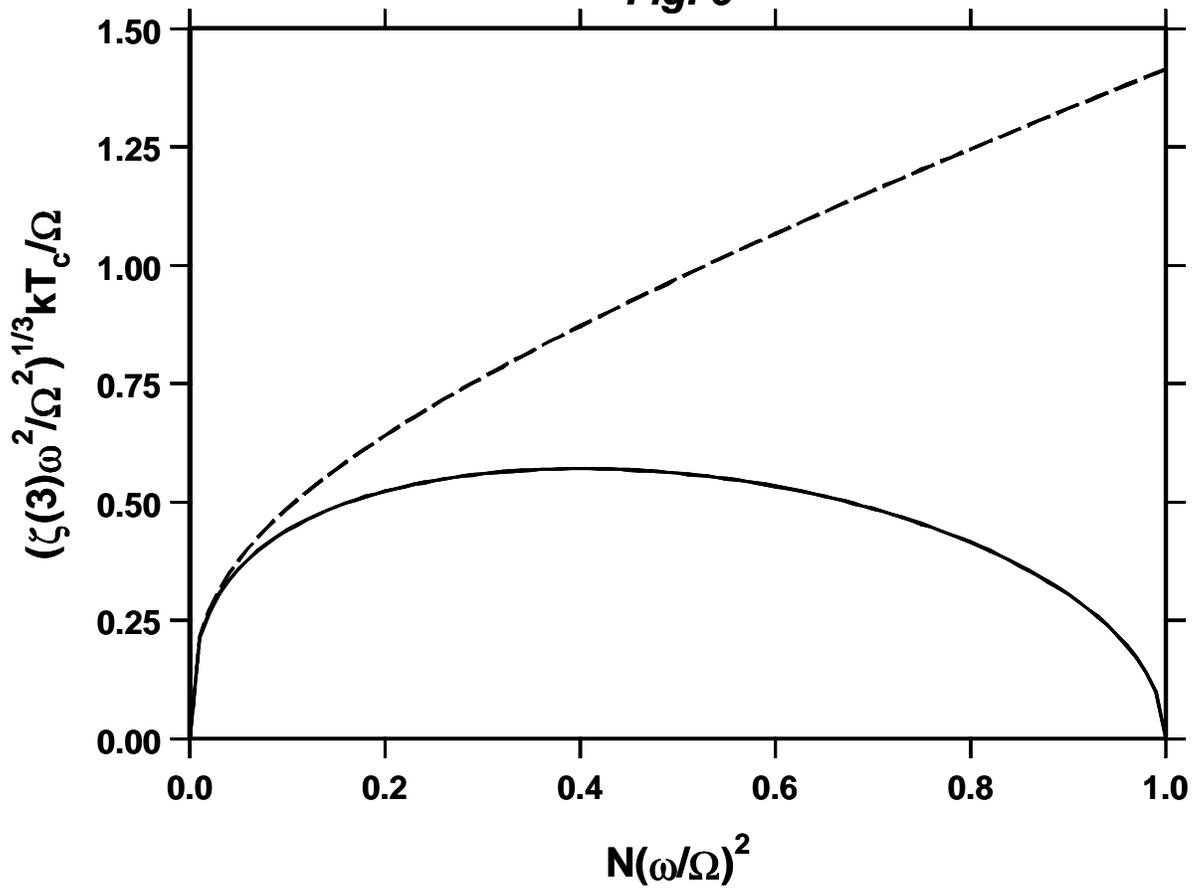
\item[Fig. 6:]  Critical temperature $T_c$ as a function of the number of
particles $N$ for a repulsive (full line) and an attractive (dashed line)
two-body interaction.

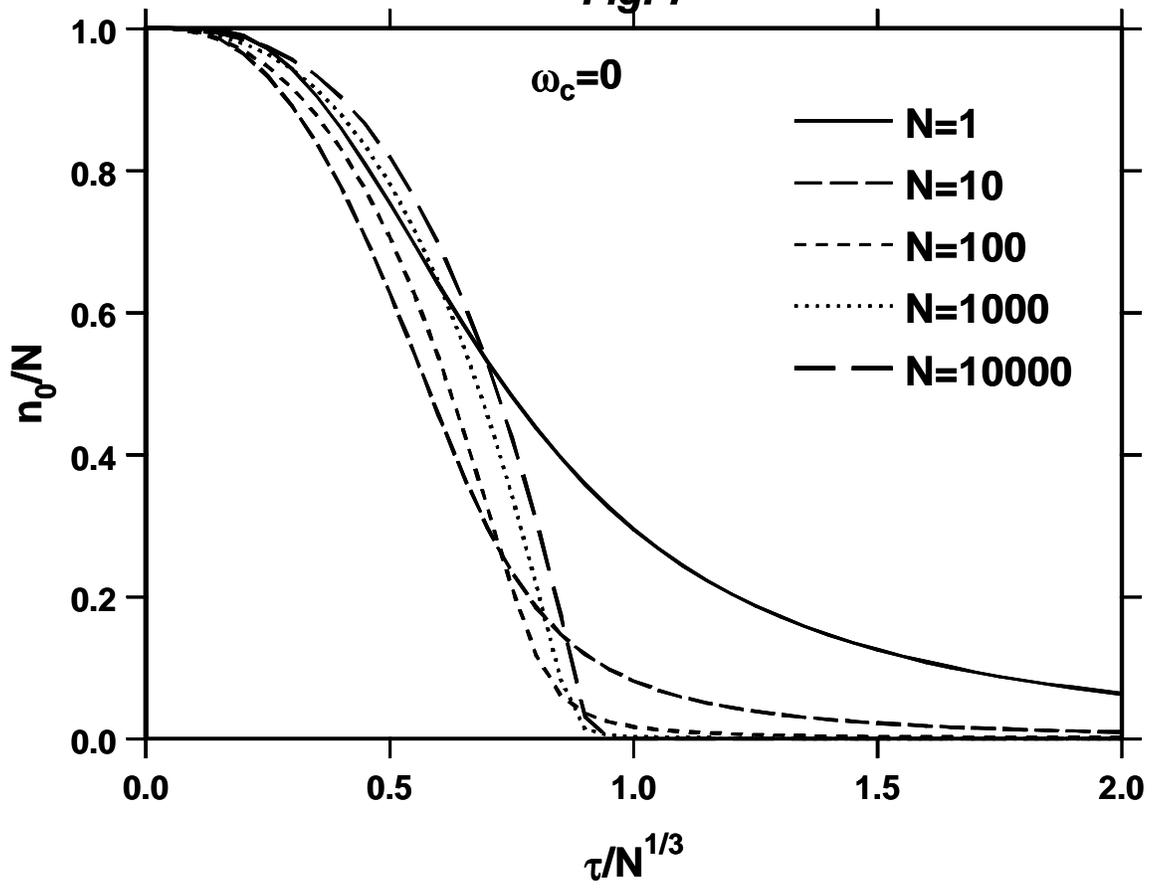
\item[Fig. 7:]  Relative ground state occupancy $n_0/N$ as a function of $%
\tau /N^{1/3}$ with $\tau $ the reduced temperature $\tau =kT/w$ for $\omega
_c=0$ and for several values of $N.$

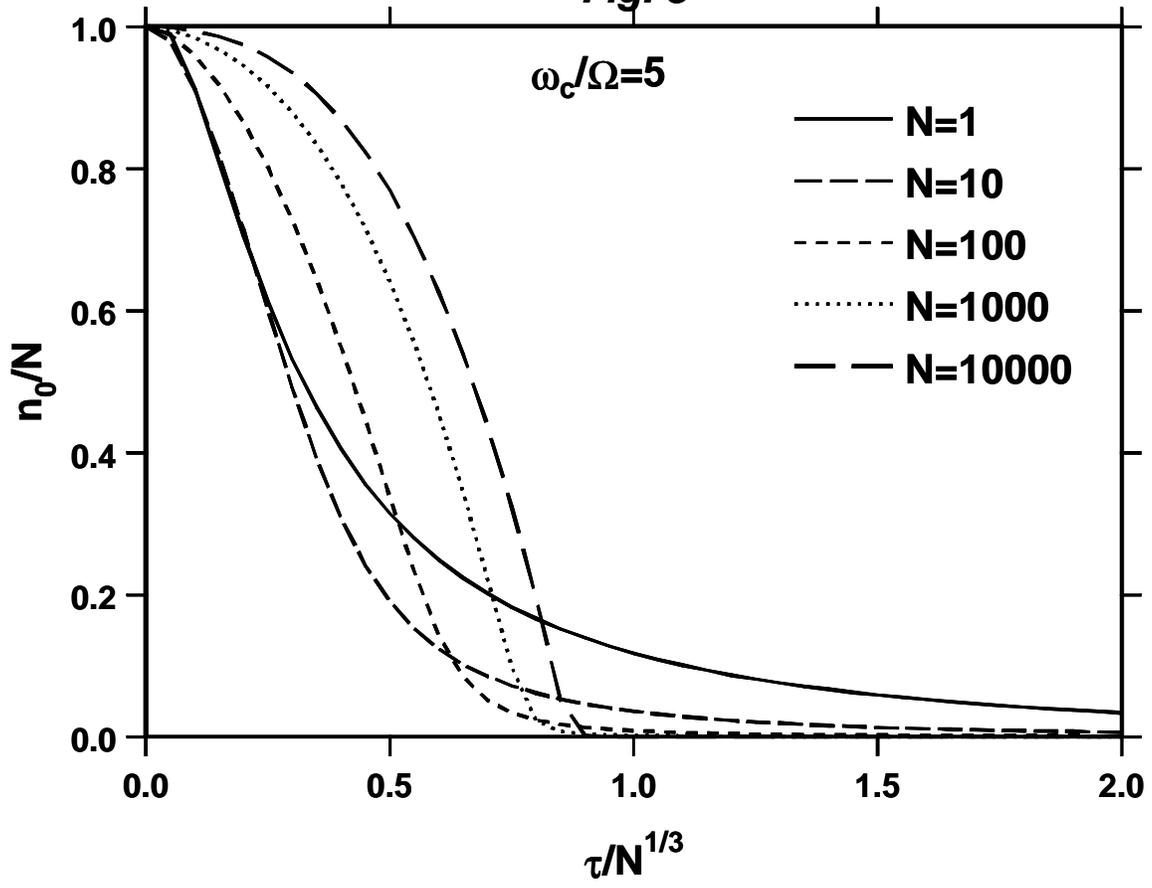
\item[Fig. 8:]  Relative ground state occupancy $n_0/N$ as a function of $%
\tau /N^{1/3}$ with $\tau $ the reduced temperature $\tau =kT/w$ for $\omega
_c/\Omega =5$ and for several values of $N$.

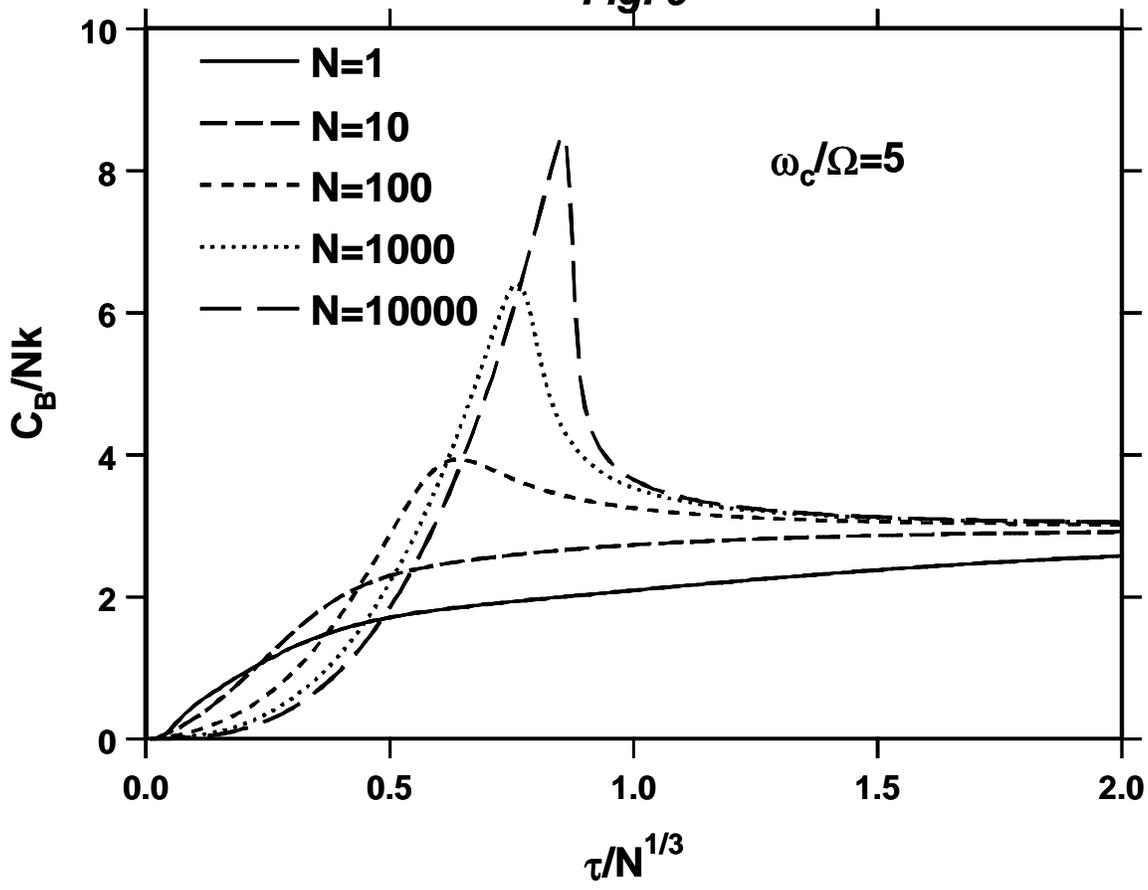
\item[Fig. 9:]  Specific heat ${\bf C}_B/Nk$ per particle in units of the
Boltzmann constant $k$ as a function of $\tau /N^{1/3}$ with $\tau $ the
reduced temperature $\tau =kT/w$ for $\omega _c/\Omega =5$ and for several
values of $N.$
\end{description}

\end{document}